\documentclass{article}
\usepackage{ijcai23}

\usepackage{times}
\usepackage{soul}
\usepackage{url}
\usepackage{hyperref}
\hypersetup{hidelinks}
\usepackage[utf8]{inputenc}
\usepackage[small]{caption}
\usepackage[labelformat=parens]{subcaption}
\usepackage{graphicx}
\usepackage{amsmath}
\usepackage{amssymb}
\usepackage{amsthm}
\usepackage{booktabs}
\usepackage{algorithm}
\usepackage{algorithmic}
\usepackage{svg}
\usepackage[switch]{lineno}
\usepackage{multirow}

\title{Graph-based Polyphonic Multitrack Music Generation}

\author{
Emanuele Cosenza$^1$
\and
Andrea Valenti$^1$\And
Davide Bacciu$^1$\\
\affiliations
$^1$University of Pisa\\
\emails
e.cosenza3@studenti.unipi.it, andrea.valenti@phd.unipi.it, davide.bacciu@unipi.it
}

\begin{document}

\maketitle

\begin{abstract}
    Graphs can be leveraged to model polyphonic multitrack symbolic music, where notes, chords and entire sections may be linked at different levels of the musical hierarchy by tonal and rhythmic relationships. Nonetheless, there is a lack of works that consider graph representations in the context of deep learning systems for music generation. This paper bridges this gap by introducing a novel graph representation for music and a deep Variational Autoencoder that generates the structure and the content of musical graphs separately, one after the other, with a hierarchical architecture that matches the structural priors of music. By separating the structure and content of musical graphs, it is possible to condition generation by specifying which instruments are played at certain times. This opens the door to a new form of human-computer interaction in the context of music co-creation. After training the model on existing MIDI datasets, the experiments show that the model is able to generate appealing short and long musical sequences and to realistically interpolate between them, producing music that is tonally and rhythmically consistent. Finally, the visualization of the embeddings shows that the model is able to organize its latent space in accordance with known musical concepts.
\end{abstract}


\section{Introduction}
The automatic generation of artistic artifacts is gathering increasing interest, also thanks to the possibilities offered by modern deep generative models.
The visual arts of painting and photography \cite{ramesh2022hierarchical}, the written expressions of prose and poetry \cite{brown2020language}, and intricate art forms such as music \cite{agostinelli2023musiclm} are all domains where neural models can be leveraged to produce realistic, if not artistically appealing, artifacts.

Despite these achievements, a closer inspection is often enough to detect whether a piece of art is the outcome of an automatic artificial process or not. While being very good at approximating the external appearance of the artworks, artificial models still lack a way to convey an artistic message to the overall experience. This results in artworks that are convincing but soulless, lacking a general coherence and a deeper meaning. This is particularly true in the case of music, where the artist needs to be very aware of the emotions evoked by a particular sequence of notes in order to stimulate a specific mood in the listener. 

A way to circumvent the above issues is to look at deep learning models as a powerful support to the human artist, instead of as a replacement. The models can thus be used as a way to automatize the low-level routine sub-tasks of the creative process, while leaving the artist free to concentrate on the overall picture. Thus, the neural network becomes an extremely versatile tool in the hand of the artist, which should be able to control and shape the output of the network in any way they see appropriate. 


In this paper, we introduce a new model for the automatic generation of symbolic sequences of multitrack, polyphonic music. The generation process is carried out through the use of a novel graph-based internal representation, which allows to explicitly model the different chords in the song and the relations between them. This representation allows the human artist to perform controlled changes to the output of the neural network in order to control specific aspects of the artistic performance, while leaving the model free to generate the remaining part in a coherent way.

The main contributions of this paper are the following:
\begin{enumerate}
\item We propose a novel graph representation of multitrack, polyphonic music, where nodes represent the chords played by different instruments and edges model the relationships between them.
\item We introduce a deep Variational Autoencoder \cite{kingma2013auto} that generates musical graphs by separating their rhythmic structure and tonal content. To the best of our knowledge, this is the first time in literature that Deep Graph Networks \cite{bacciu2020gentle} are used to generate multitrack, polyphonic music. 
\item We show a new generative scenario enabled by our approach in which the user can intuitively condition generation by specifying which instruments have to be played at specific timesteps.
\end{enumerate}

\section{Related Works}

In recent years, there have been many attempts at generating symbolic music with deep learning. Various works have focused on sequential models such as LSTMs \cite{chu2016song,brunner2017jambot,roberts2018hierarchical} and, more recently, Transformers \cite{huang2018music,valenti2021calliope}. When considering specific representations (e.g. pianorolls), music can also be processed by convolutional networks, with associated applications to music generation \cite{chuan2018modeling,huang2019counterpoint}.

Variational Autoencoders (VAE) \cite{kingma2013auto} and Generative Adversarial Networks (GAN) \cite{goodfellow2014generative} have emerged as plausible candidates for symbolic music generation. MidiNet \cite{yang2017midinet}, C-rnn-gan \cite{mogren2016c} and MuseGAN \cite{dong2018musegan} are all models in which a convolutional or recurrent generator produces music from a random sample, 
and a discriminator is trained to distinguish generated samples from real ones. For what concerns VAEs, an early approach to music generation is Midi-VAE \cite{brunner2018midi}, where separate GRU encoder/decoder pairs are used for pitch, instrument and velocity, while sharing the same latent space. In \cite{roberts2018hierarchical}, instead, a high-level conductor LSTM takes the latent code generated by an encoder and produces latent variables corresponding to different segments of music. These are then processed by a lower-level decoder LSTM, which focuses on the generation of smaller subsections one note at a time. The same hierarchical approach is followed by PianoTree VAE \cite{wang2020pianotree}, which uses multiple GRUs to compute bar decodings from a latent code representing the entire piece, and chord decodings from bar decodings. The authors also exploit note-chord hierarchy priors, computing chord embeddings from note embeddings. An interesting middle ground between VAEs and GANs is represented by Adversarial Autoencoders (AAE) \cite{makhzani2015adversarial}, which have been used in the context of music generation to impose arbitrary priors to latent variables \cite{valenti2020learning,valenti2021calliope}.\\

A challenge in devising symbolic generators is choosing an appropriate representation for music data. Researchers have therefore started to experiment with graph-based representations, where musical entities and their relationships are modeled, respectively, by nodes and edges. Musical graphs have been built at the note level \cite{liu2010complex,ferretti2018complex,ferretti2017modeling,mellon2014genre}, associating nodes to notes and edges to temporal or tonal relationships, as well as at a higher level of the hierarchy, using melodic segments \cite{simonetta2018symbolic} and bars \cite{wu2020popmnet,zou2021melons} as building blocks.

In the literature, there is a substantial lack of studies that consider graph representations in the context of deep learning for symbolic music. The VAE-based performance renderer in \cite{wu2020popmnet} and the cadence detector in \cite{karystinaios2022cadence} are, to the best of our knowledge, the only systems that use Deep Graph Networks to process musical graphs. In both works, graphs are constructed at the note level and edges represent both tonal and temporal relationships between musical entities. For what concerns generation, the only attempts at using graphs with deep learning are represented by PopMNet \cite{wu2020popmnet} and MELONS \cite{zou2021melons}. Both works use GANs and recurrent networks, enforcing the typical structure of human music through a bar-level graph representation. These graphs are used to condition the generation of monophonic music, which is carried out by the recurrent networks. In contrast to these works, our approach uses graphs at a lower level, leveraging Deep Graph Networks to automatically learn meaningful tonal and rhythmic concepts in the context of polyphonic multitrack music generation.

\section{Graph-based Music Generation}

The proposed model processes polyphonic, multitrack music. Input songs are assumed to be available as an $N \times I \times T \times P$ multitrack pianoroll binary tensor, where $N$ is the number of bars, $I$ the number of tracks, $T$ the number of timesteps in a bar and $P$ the number of possible pitches. An example of a multitrack pianoroll is shown in Figure \ref{fig:pianoroll}. The number of timesteps in a bar, $T$, is fixed to 32, which allows to represent notes with rhythmic value $1/32$. A note is defined by its pitch and duration values. Songs are assumed to contain a set of tracks played by non-percussive instruments together with a drum/percussion track (possibly silenced). This can be easily enforced during the preprocessing phase.

\subsection{Graph-based Music Representation}

\begin{figure}[t]
    \centering
    \begin{subfigure}{0.12\textwidth}
        \centering
        \includegraphics[width=1\textwidth]{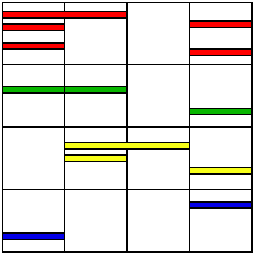}
        \caption{}
        \label{fig:pianoroll}
    \end{subfigure}
    \hfil
    \begin{subfigure}{0.12\textwidth}
        \centering
        \includegraphics[width=1\textwidth]{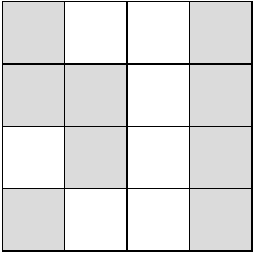}
        \caption{}
        \label{fig:activations}
    \end{subfigure}
    \par\bigskip
    \begin{subfigure}{0.13\textwidth}
        \centering
        \includegraphics[width=1\textwidth]{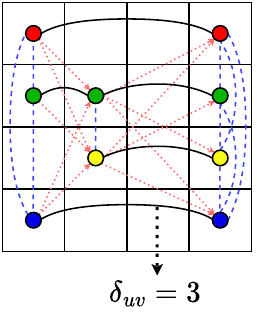}
        \caption{}
        \label{fig:clg}
    \end{subfigure}
\label{fig:graph}
\caption{(a) Illustration of a single bar of a multitrack pianoroll with four tracks (rows) and four timesteps (columns). Colored rectangles in the grid represent the notes being played in the sequence, while their position inside each cell indicates their pitch. (b) A structure tensor computed from the pianoroll. (c) The resulting chord level graph. In the image, black, solid connections indicate track edges. Red, dotted connections indicate next edges. Blue, dashed connections indicate onset edges. Edge features $\delta_{uv}$ indicate the distance in timesteps between two nodes. The content of the graph is omitted for simplicity.}
\end{figure}

We propose to represent polyphonic multitrack music by a \textit{chord-level graph} $g=(\mathcal{V}, \mathcal{E}, \mathcal{A}, \mathcal{X})$, where $\mathcal{V}$ is the set of nodes, $\mathcal{E}$ is the set of (multi-type) edges, $\mathcal{A}$ the set of edge features and $\mathcal{X}$ the set of node features. An example of a chord-level graph is shown in Figure \ref{fig:clg}.

The \textit{structure} $\mathcal{S}$ of $g$ is represented by the sets $\mathcal{V}$, $\mathcal{A}$ and $\mathcal{E}$. Each node $v \in \mathcal{V}$ corresponds to the activation of a chord in a specific track and timestep. We identify three types of edges $(u, v) \in \mathcal{E}$: \textit{track} edges, \textit{onset} edges and \textit{next} edges. Track edges connect nodes that represent consecutive activations of a single track. Onset edges connect nodes that represent simultaneous activations of different tracks. Finally, next edges connect nodes that represent consecutive activations of different tracks in different timesteps. In order to model different tracks, a separate track edge type is instantiated for each track. Track edges model \textit{intra-track} relationships since they only connect nodes belonging to a single track. On the other hand, onset and next edges model \textit{inter-track} relationships since they connect nodes related to different tracks. Each edge feature $a_{uv} \in \mathcal{A}$ contains the type of the edge $(u, v)$ as well as the distance in timesteps between the two nodes.

The \textit{content} $\mathcal{C}$ of $g$ is represented by the set of node features $\mathcal{X}$. Node features $x_v \in \mathcal{X}$ contain the list of notes played in correspondence of node $v$. The number of maximum notes in a chord, $\Sigma$, is fixed a priori. Each note is represented as a feature vector of dimension $D$. The vector contains information about pitch and duration stored as a one-hot token pair. The pitch token can assume 131 different values, which correspond to 128 MIDI pitches with the addition of $\texttt{SOS}_P$, $\texttt{EOS}_P$ and $\texttt{PAD}_P$ tokens. Similarly, the duration token can assume 99 different values, which correspond to 96 different durations (yielding a maximum duration of 3 bars) with the addition of $\texttt{SOS}_D$, $\texttt{EOS}_D$ and $\texttt{PAD}_D$ tokens.

The structure of $g$ is encoded by the tensor $\bold{S} \in \{0, 1\}^{N \times I \times T}$, where $\bold{S}_{n, i, t} = 1$ if and only if there is an activation of at least one note in the track $i$ at timestep $t$ of the $n$-th bar. Intuitively, $\bold{S}_{n, i, t}$ indicates whether track $i$ is active (not counting the sustain of notes) at timestep $t$ in the $n$-th bar. An example of a structure tensor is shown in Figure \ref{fig:activations}. The content of a chord-level graph, on the other hand, can be encoded through a tensor $\bold{X} \in \mathbb{R}^{|\mathcal{V}| \times \Sigma \times D}$ after fixing an ordering of $\mathcal{V}$.

\subsection{Deep Graph Network for Music}

Our graph-based representation of music is processed by a deep VAE \cite{kingma2013auto} that reconstructs the structure $\mathcal{S}$ and the content $\mathcal{C}$ of a chord-level graph $g=(\mathcal{S}, \mathcal{C})$. Its encoder models the encoding distribution $q_{\boldsymbol{\phi}}(\boldsymbol{z} | \mathcal{S}, \mathcal{C})$, where $\boldsymbol{z} \in \mathbb{R}^d$. The decoder network, on the other hand, models $p_{\boldsymbol{\theta}}(\mathcal{S}, \mathcal{C} | \boldsymbol{z})$. After introducing the latent variables $\boldsymbol{z}_\mathcal{S} \in \mathbb{R}^d$ and $\boldsymbol{z}_\mathcal{C} \in \mathbb{R}^d$, the generative process can be formalized as follows:
\begin{equation}
\label{eq:genmodel}
    \resizebox{.91\linewidth}{!}{$
            \displaystyle
            p_{\boldsymbol{\theta}}(\mathcal{S}, \mathcal{C}, \boldsymbol{z}_\mathcal{S}, \boldsymbol{z}_\mathcal{C} | \boldsymbol{z}) = p_{\boldsymbol{\theta}}(\boldsymbol{z}_\mathcal{S} | \boldsymbol{z}) p_{\boldsymbol{\theta}}(\boldsymbol{z}_\mathcal{C} | \boldsymbol{z}) p_{\boldsymbol{\theta}}(\mathcal{S} | \boldsymbol{z}_\mathcal{S}) p_{\boldsymbol{\theta}}(\mathcal{C} | \boldsymbol{z}_\mathcal{C}, \mathcal{S})
    $}
\end{equation}

\begin{figure}[b]
         \centering
         \includegraphics[width=0.45\textwidth]{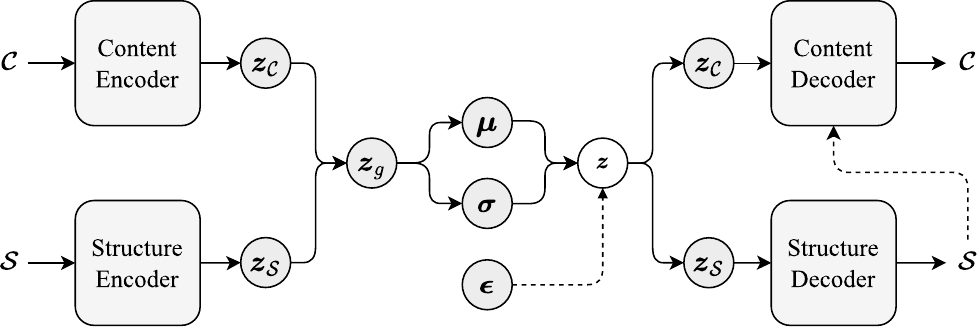}
\caption{High-level visualization of the model.\label{fig:model}}
\end{figure}

A high-level representation of the model is shown in Figure \ref{fig:model}.
The encoder consists of two separate submodules, namely a \textit{content encoder} and a \textit{structure encoder} which output, respectively, the codes $\boldsymbol{z}_\mathcal{S}$ and $\boldsymbol{z}_\mathcal{C}$. The two codes are finally combined into  a graph code $\boldsymbol{z}_g$ with a linear layer. The decoder, on the other hand, generates the structure $\mathcal{S}$ and the content $\mathcal{C}$ of $g$ one after the other. First, symmetrically to the encoder, it decomposes $\boldsymbol{z}$ into two separate latent vectors $\boldsymbol{z}_\mathcal{S}$ and $\boldsymbol{z}_\mathcal{C}$ through a linear layer. Then, it generates $\mathcal{S}$ from $\boldsymbol{z}_\mathcal{S}$ through a \textit{structure decoder} and the content $\mathcal{C}$ from $\mathcal{S}$ and $\boldsymbol{z}_\mathcal{C}$ through a deep graph \textit{content decoder}. The content and the structure decoder model, respectively, the distributions $p_{\boldsymbol{\theta}}(\mathcal{S}|\boldsymbol{z})$ and $p_{\boldsymbol{\theta}}(\mathcal{C}|\mathcal{S}, \boldsymbol{z})$.

\paragraph{Content Encoder.}

\begin{figure*}
    \centering
    \begin{subfigure}{0.45\textwidth}
        \centering
        \includegraphics[width=1\textwidth]{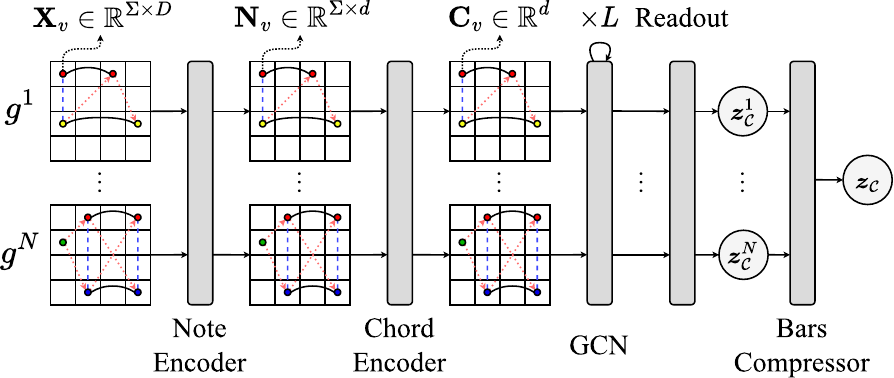}
        \caption{}
        \label{fig:contentenc}
    \end{subfigure}
    \hspace{1 cm}
    \begin{subfigure}{0.45\textwidth}
        \centering
        \includegraphics[width=1\textwidth]{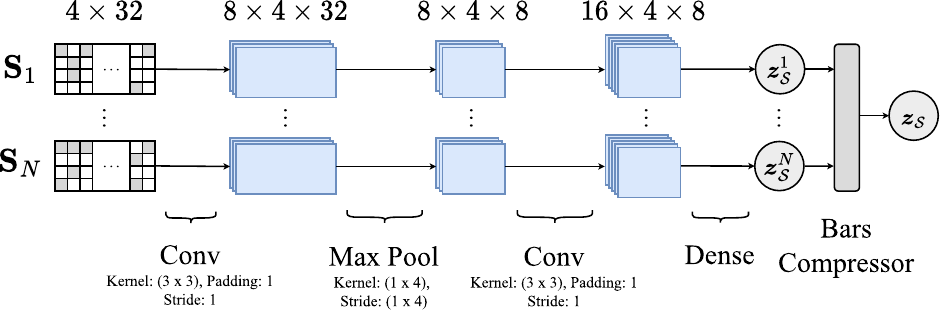}
        \caption{\label{fig:structureenc}}
    \end{subfigure}
\caption{Visualization of the content encoder (\subref{fig:contentenc}) and the structure encoder (\subref{fig:structureenc}).}
\end{figure*}

The content encoder (Figure \ref{fig:contentenc}) develops progressively higher-level representations for notes, chords, bars and the whole piece. 
This module first embeds each note in a $d$-dimensional space with a note encoder, which uses separate embedding matrices for pitches and durations. Next, a chord encoder processes the list of notes associated to each node, producing $d$-dimensional chord representations. In our instantiation of the model, the chord encoder is implemented as a linear layer that takes a concatenation of the $\Sigma$ note representations and yields a final chord embedding. These chord representations are the initial node states $\boldsymbol{h}_v^0$ of an encoder Graph Convolutional Network (GCN) \cite{bacciu2020gentle} with $L$ layers. Combining the techniques employed in \cite{schlichtkrull2018modeling,simonovsky2017dynamic,gilmer2017neural}, the GCN constructs new node states by taking into account the discrete information regarding both the edge types and the distances between nodes. Residual connections are used between consecutive layers in the GCN. This has proven to be beneficial in mitigating oversmoothing problems with large values of $L$ \cite{li2018deeper,li2019deepgcns}. Batch Normalization \cite{ioffe2015batch} is used after each graph convolutional layer to speed up convergence and improve the generalization capability of the model. We refer the reader to the supplementary material\footnote{\url{https://emanuelecosenza.github.io/polyphemus/assets/suppmaterials.pdf}} for details about the implementation of the GCN.

After $L$ graph convolutional layers, a readout layer aggregates the information contained in each subgraph $g^n$ of $g$ related to the $n$-th bar of the musical sequence. This layer, which resembles the ones in \cite{jeong2019graph} and \cite{li2015gated}, produces bar embeddings $\boldsymbol{z}_\mathcal{C}^1, \dots, \boldsymbol{z}_\mathcal{C}^N$ using a soft attention layer, which is in charge of learning the importance of single track activations.

The $N$ bar embeddings $\boldsymbol{z}_\mathcal{C}^1, \dots, \boldsymbol{z}_\mathcal{C}^N$ are concatenated and passed through a bar compressor, which is implemented as a linear layer, to obtain the final content representation $\boldsymbol{z}_{\mathcal{C}}$.

\paragraph{Structure Encoder.}  

The structure encoder (Figure \ref{fig:structureenc}) takes as input the structure tensor $\bold{S} \in \mathbb{R}^{N \times I \times T}$ and computes the code $\boldsymbol{z}_\mathcal{S}$. This module first encodes each bar $\bold{S}_n \in \mathbb{R}^{I \times T}$ into a latent representation $\boldsymbol{z}_\mathcal{S}^n \in \mathbb{R}^d$ through a CNN \cite{Goodfellow-et-al-2016} made of two convolutional layers with ReLU activations and Batch Normalization, interleaved by max pooling.
The bar representations $\boldsymbol{z}_\mathcal{S}^1, \dots, \boldsymbol{z}_\mathcal{S}^N$ are then computed by passing the signal through two dense layers. These representations are finally concatenated and passed through a linear layer to obtain $\boldsymbol{z}_\mathcal{S}$.

\paragraph{Structure Decoder.}

\begin{figure*}
         \centering
         \begin{subfigure}{0.4\textwidth}
         \centering
         \includegraphics[width=1\textwidth]{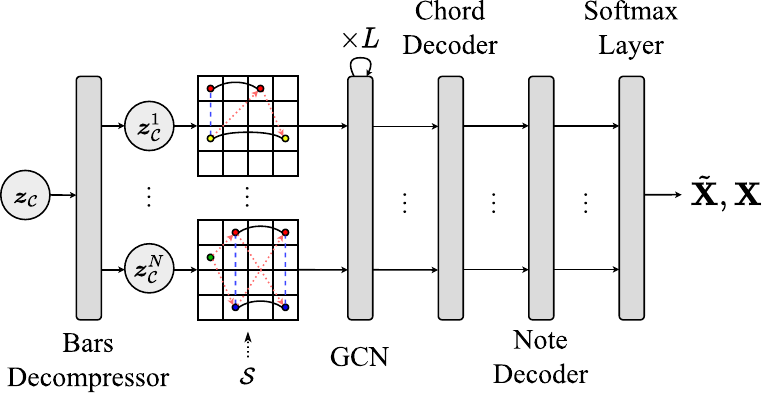}
         \caption{\label{fig:contentdec}}
         \end{subfigure}
         \hspace{1 cm}
         \begin{subfigure}{0.45\textwidth}
            \centering
            \includegraphics[width=1\textwidth]{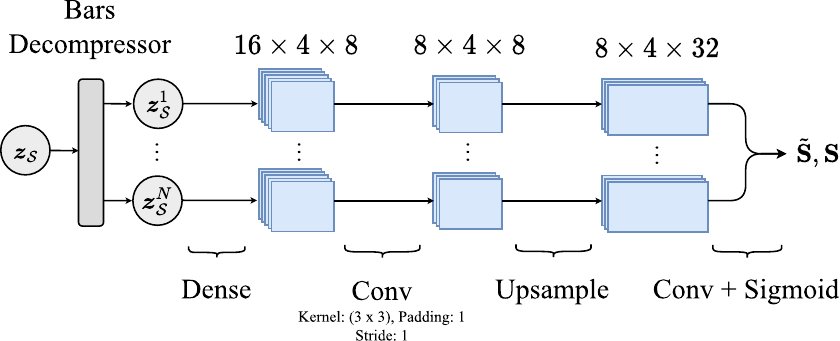}
         \caption{\label{fig:structdec}}
         \end{subfigure}
         
\caption{Visualization of the content decoder (\subref{fig:contentdec}) and the structure decoder (\subref{fig:structdec}).}
\end{figure*}

The structure decoder (see Figure \ref{fig:structdec}) is specular to the structure encoder. It first decompresses $\boldsymbol{z}_\mathcal{S}$ into $N$ structure bar representations $\boldsymbol{z}_\mathcal{S}^1, \dots, \boldsymbol{z}_\mathcal{S}^N$ and decodes each of them into a structure tensor $\bold{S}_n \in \mathbb{R}^{I \times T}$ with a bar decoder. The bar decoder mirrors the bar encoder, with the difference that upsample layers are interleaved with convolutional layers to obtain the original resolution of the pianoroll. Finally, a sigmoid layer produces probability values which are stacked to form the probabilistic structure tensor $\bold{\Tilde{S}}$.

\paragraph{Content Decoder.}
It reconstructs the content of $g$ from $\boldsymbol{z}_\mathcal{C}$ and $\mathcal{S}$. The decoder first decompresses $\boldsymbol{z}_\mathcal{C}$ into $\boldsymbol{z}_\mathcal{C}^1, \dots, \boldsymbol{z}_\mathcal{C}^N \in \mathbb{R}^d$. Each $\boldsymbol{z}_\mathcal{C}^n$ is used to initialize the states of the nodes in the subgraph $g^n$, which represents the connected component related to the $n$-th bar of the structure $\mathcal{S}$. From there, a GCN identical to the one employed in the encoder computes the final states $\bold{h}_v^L \in \mathbb{R}^d$ for each node $v$. At this point, a (linear) chord decoder transforms each final node state $\bold{h}_v^L$ into the corresponding $\Sigma$ note representations of dimension $d$. Such note representation vectors are split into two halves: each half is transformed by a pitch and a duration decoder, respectively, into pitch and duration information. As in the encoder, two separate pitch embedding matrices are used for drum and non-drum pitches. Finally, a softmax layer outputs two separate probability distributions over pitches and durations, yielding the probabilistic tensors $\bold{\Tilde{P}}$ and $\bold{\Tilde{D}}$, which contain, respectively, pitch and duration probabilities.

\subsection{Training}

The model is trained to minimize the following loss
\begin{equation}
\label{eq:betavae}
\resizebox{.91\linewidth}{!}{$
            \displaystyle
            \mathcal{L}(g) =  \mathbb{E}[-\log p_{\boldsymbol{\theta}}(g|\boldsymbol{z})] + \beta D_{KL}(q_{\phi}(\boldsymbol{z}|g) \, || \, \mathcal{N}(\boldsymbol{0}, \boldsymbol{I})),
        $}
\end{equation}
where $D_{KL}(\cdot||\cdot)$ is the KL divergence and the expectation is taken with respect to $\boldsymbol{z} \sim q_{\boldsymbol{\phi}}(\boldsymbol{z}|\boldsymbol{x})$. Following the $\beta$-VAE framework \cite{higgins2016beta}, the hyperparameter $\beta$ controls the trade-off between reconstruction accuracy and latent space regularization.

Since the generative process is divided in two parts, the log-likelihood term in Equation \ref{eq:betavae} can be decomposed as follows: 
\begin{equation} \label{eq:decomp}
\begin{split}
\log p_{\boldsymbol{\theta}}(g|\boldsymbol{z}) & = \log \left( p_{\boldsymbol{\theta}}(\mathcal{S}|\boldsymbol{z}) p_{\boldsymbol{\theta}}(\mathcal{C}|\boldsymbol{z}, \mathcal{S}) \right) \\
& = \log p_{\boldsymbol{\theta}}(\mathcal{S}|\boldsymbol{z}) + \log p_{\boldsymbol{\theta}}(\mathcal{C}|\boldsymbol{z}, \mathcal{S}).
\end{split}
\end{equation}

The first term in Equation \ref{eq:decomp} can be derived in the following way:
\begin{equation}
\begin{split}
\log p_{\boldsymbol{\theta}}(\mathcal{S}|\boldsymbol{z})
    & = \sum_{n, i, t} \bold{S}_{n, i, t} \log \bold{\Tilde{S}}_{n, i, t} + \\ & + (1-\bold{S}_{n, i, t}) \log (1-\bold{\Tilde{S}}_{n, i, t}),
\end{split}
\end{equation}
where independence is assumed between variables.

Computing the content log-likelihood in Equation \ref{eq:decomp} is trickier, since the structure generated by the structure decoder may be different from the real one. We circumvent this problem by using a form of teacher forcing, where the content is obtained by filling the real structure in place of the one generated by the structure decoder. In this way, the following likelihood can always be computed:
\begin{equation}
\label{eq:contentloglike}
    \begin{split}
    \log p_{\boldsymbol{\theta}}(\mathcal{C}|\boldsymbol{z}, \mathcal{S}) & = \sum_{i} \sum_{\sigma} \log (\bold{\Tilde{P}}_{i, \sigma})^T \bold{P}_{i, \sigma} + \\ & + \log (\bold{\Tilde{D}}_{i, \sigma})^T \bold{D}_{i, \sigma}, \end{split}
\end{equation}
where $\bold{P}$ and $\bold{D}$ are tensors containing, respectively, real one-hot pitch and duration tokens, while $\bold{\Tilde{P}}$ and $\bold{\Tilde{D}}$ represent their probabilistic reconstructions. Indepedence is assumed between all pitch and duration variables.

\section{Experiments}

Following \cite{roberts2018hierarchical,valenti2020learning,valenti2021calliope}, we experiment on short and long sequences of MIDI music. The experiments probe the generative capabilities of the model comparing, whenever possible, to state of the art approaches. 
We further examine a novel scenario enabled by our methodology where generation is conditioned on user-specified structures. Finally, pitch, duration and chord embeddings are visualized to show that the model is able to learn known tonal and rhythmic concepts. We refer the reader to the source code\footnote{\url{https://github.com/EmanueleCosenza/polyphemus}} and the additional material\footnote{\url{https://emanuelecosenza.github.io/polyphemus/}}, which contains the audio samples produced in the experimental phase. 

\subsection{Data and Experimental Setup}

We consider the `LMD-matched' version of the Lakh MIDI Dataset \cite{raffel2016learning}, which contains a total of 45,129 MIDI songs scraped from the internet. Additionally, we train our models on the more challenging MetaMIDI Dataset (MMD) \cite{ens2021building}, a recent and unexplored large scale MIDI collection totalling 436,631 songs. For each dataset, we obtain two new datasets containing, respectively, 2-bar and 16-bar sequences represented as chord-level graphs. The preprocessing pipeline is similar to that in \cite{roberts2018hierarchical,valenti2020learning,valenti2021calliope}. The details about preprocessing can be found in the supplementary material. At the end of this phase, each sequence is composed of 4 tracks: a drum track, a bass track, a guitar/piano track and a strings track. The sizes of the resulting datasets are shown in Table \ref{tab:datasets}.

\begin{table}
    \centering
    \begin{tabular}{lrr}
        \toprule
        Datasets & 2 Bars & 16 Bars \\
        \midrule
        LMD-matched     & 6,813,946 & 2,842,739        \\
        MetaMIDI Dataset  & 11,076,635          & 27,251,322        \\
        \bottomrule
    \end{tabular}
    \caption{Size of the datasets obtained in the preprocessing phase.}
    \label{tab:datasets}
\end{table}

The experiments focus on two versions of the model, one for 2-bar sequences and one for 16-bar sequences. We use for both a 70/10/20 split. The number of layers $L$ of the encoder and decoder GCN is fixed to 8. The value $d$ is set to 512. Adam \cite{kingma2014adam} is used as the optimizer for both models. The initial learning rates are set to 1e-4 and 5e-5, respectively, for the 2-bar and the 16-bar model. In both cases, the learning rate is decayed exponentially after 8000 gradient updates with a decay factor of $1-5\text{e-}6$. The hyperparameter $\beta$ is annealed from 0 to 0.01. In the first 40,000 gradient updates, $\beta$ is always 0, allowing the model to focus on the reconstruction task to find good initial representations. After this phase, the hyperparameter is annealed every $u=40,000$ gradient updates by adding 0.001 to its current value. The batch size $b$ is set to 256 and 32 respectively for the 2-bar and the 16-bar model.

\subsection{Generation}

\begin{figure}[b]
    \centering
    \includegraphics[width=1\columnwidth]{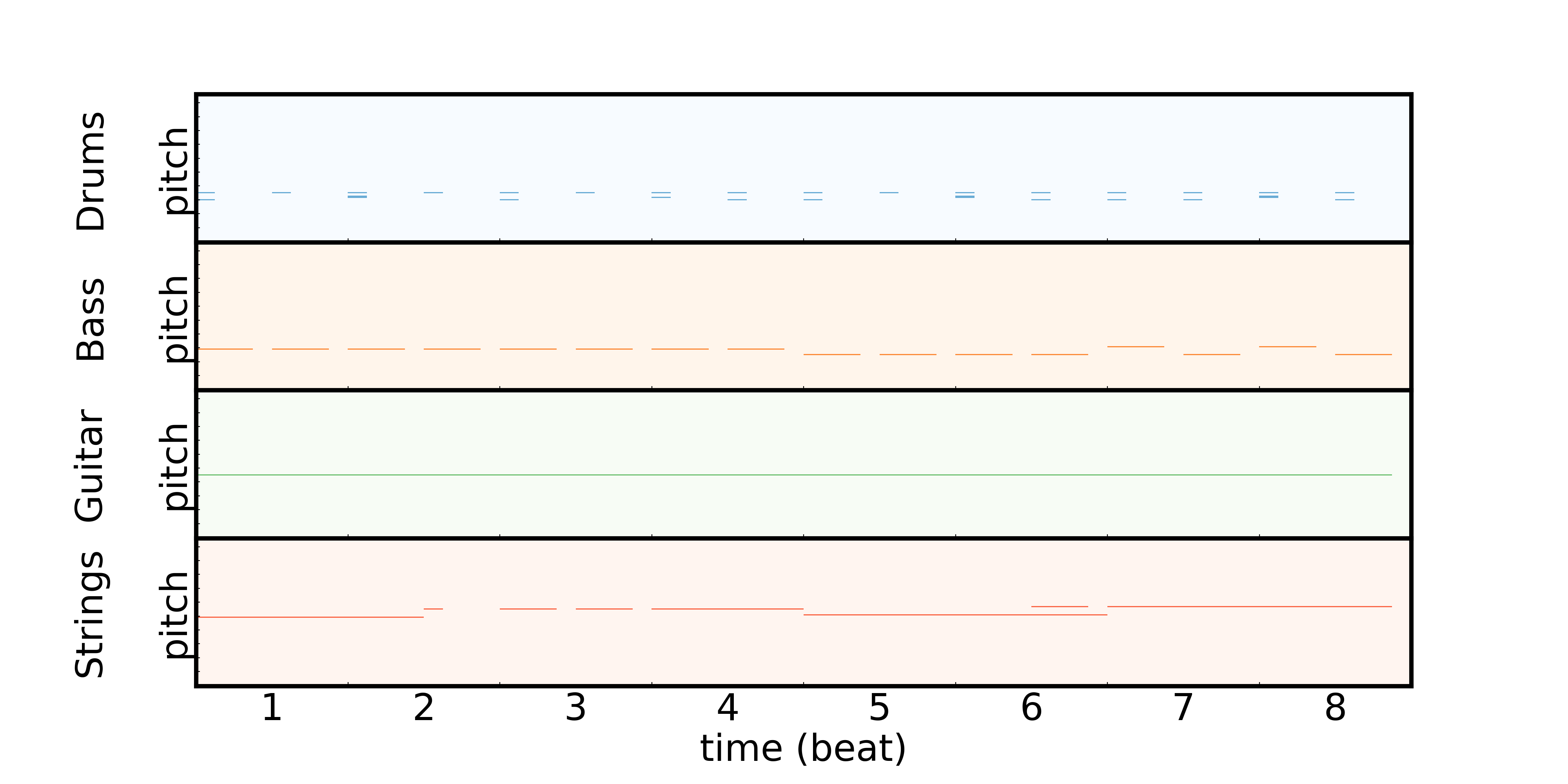}
    \caption{A pianoroll of a generated 2-bar sequence.\label{fig:gen2}}
\end{figure}

The first set of experiments concerns the analysis of sequences generated from random codes $\boldsymbol{z}$. A qualitative visual inspection of the samples suggests that the models can consistently generate realistic music. Figure \ref{fig:gen2} shows an example of a 2-bar generated sequence. As can be seen from the pianoroll, the generated structures are well organized rhythmically, with drum and bass events played at the same timesteps. In the listening analysis\footnote{Audio samples of generated 2-bar and 16-bar samples can be found here: \url{https://emanuelecosenza.github.io/polyphemus/generation.html}}, the 2-bar models appear to be particularly consistent, producing reasonable chord progressions, melodic segments and drum patterns. The 16-bar models are also coherent, both rhythmically and tonally. However, 16-bar sequences generally lack variability as the models tend to repeat musical structures across bars with slight differences. The rhythmic consistency of the model is enforced by the fact that the content decoder can focus on the generation of reasonable rhythmic patterns. The tonal consistency, instead, is ensured by the expressiveness of the GCN decoder, which is able to fill chord-level graphs realistically.
To provide a more quantitative assessment, following previous works \cite{dong2018musegan,valenti2021calliope}, we measure the generative ability of the trained models by computing the following metrics on 20,000 generated sequences: 
\begin{itemize}
    \item EB (Empty Bars): ratio of empty bars.
    \item UPC (Used Pitch Classes): number of used pitch classes (12) per bar.
    \item DP (Drum Patterns): ratio of notes in 16-beat patterns, which are common in popular music (in \%).
\end{itemize}
\begin{table*}
    \centering
    \resizebox{0.7\textwidth}{!}{%
    \begin{tabular}{clrrrrrrrr}
    \toprule
    {} & {} & \multicolumn{4}{c}{EB} & \multicolumn{3}{c}{UPC $\downarrow$} & \multicolumn{1}{c}{DP $\uparrow$} \\ 
    \cmidrule(lr){3-6} \cmidrule(lr){7-9} \cmidrule(lr){10-10}
    {} & {} & D & B & G/P & S & B & G & S & D \\ 
    \midrule
    \multirow{6}{*}{LMD-matched} & jamming & 6.59 & 2.33 & 20.45 & 6.10 & \textbf{1.53} & 3.91 & 4.09 & 93.2 \\
    {} & composer & 0.01 & 28.9 & 1.35 & 0.01 & 2.51 & 4.55 & 5.19 & 75.3 \\
    {} & hybrid & 2.14 & 29.7 & 14.75 & 6.04 & 2.35 & 5.11 & 5.24 & 71.3 \\
    \cmidrule{2-10}
    {} & Calliope & 0.0 & 0.0 & 0.0 & 0.0 & 2.08 & 3.87 & 2.52 & 94.84 \\
    \cmidrule{2-10}
    {} & \textbf{Ours (2-bars)} & 4.58 & 4.39 & 20.46 & 17.74 & 2.27 & 2.53 & 2.72 & \textbf{96.97} \\
    {} & \textbf{Ours (16-bars)} & 1.96 & 3.37 & 11.38 & 12.02 & 1.79 & \textbf{2.38} & \textbf{2.07} & 96.59 \\
    \midrule
    \multirow{2}{*}{MetaMIDI Dataset} & \textbf{Ours (2-bars)} & 5.38 & 8.31 & 23.49 & 21.54 & 1.85 & 2.03 & 2.24 & 96.28\\
    {} & \textbf{Ours (16-bars)} & 4.20 & 7.20 & 18.39 & 17.56 & 1.34 & 1.66 & 1.39 & 95.92\\
    \bottomrule
    \end{tabular}%
    }
\caption{Generation metrics of the proposed model, Calliope and the jamming, composer and hybrid versions of MuseGAN (EB: empty bars ($\%$), UPC: number of used pitch classes, DP: drum patterns ($\%$), D: drums, B: bass, G/P: guitar/piano, S: strings).}
\label{table:genmetrics}
\end{table*}

Table \ref{table:genmetrics} shows the results obtained by our models, comparing our approach to different versions of MuseGAN \cite{dong2018musegan} and Calliope \cite{valenti2021calliope}. We also include metrics for the models trained on MetaMIDI Dataset with the goal of stimulating research on larger MIDI collections. The EB values are never equal to zero, which indicates that there are no issues with holes in the latent space and that the models do not ignore the latent codes during decoding. The UPC values are consistently low, indicating that the models have learned to stick to specific tonalities in the context of single bars. Additionally, the DP values for the proposed model are the highest, confirming its consistency on the rhythmic level. These results further validate the proposed methodology and confirm the rhythmic and tonal coherence of the model.

To inspect the structure of the latent space learned by the 2-bar Lahk MIDI Dataset model, we interpolate random latent codes linearly and we examine the music obtained by concatenating the resulting 2-bar sequences\footnote{Audio samples and pianorolls of interpolations can be found at the following link: \url{https://emanuelecosenza.github.io/polyphemus/interpolation.html}}. In the majority of cases, the interpolations created with the model are smooth and remain coherent, both tonally and rhythmically, throughout their entirety. Moreover, when the starting samples differ substantially, the model manages to create appealing transitions between distant styles. This suggests that the model has learned to organize its latent space in accordance with known musical semantics.
%
%

\subsection{Structure-conditioned Generation}

\begin{figure}[h]
    \centering
    \begin{subfigure}{0.4\textwidth}
        \centering
        \includegraphics[width=1\textwidth]{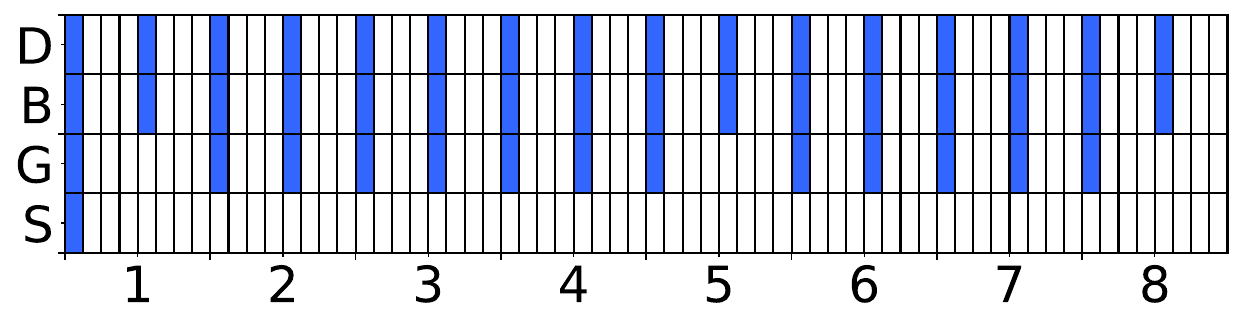}
        \caption{\label{fig:genstruct}}
    \end{subfigure}
    \par\bigskip
    \begin{subfigure}{0.4\textwidth}
        \centering
        \includegraphics[width=1\textwidth]{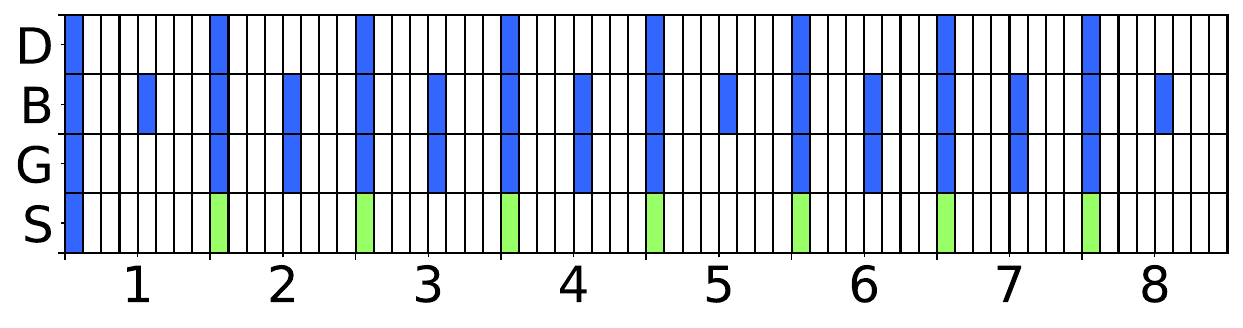}
        \caption{\label{fig:editedstruct}}
    \end{subfigure}
         
\caption{A visualization of the structure editing process. (\subref{fig:genstruct}) A structure tensor $\mathbf{S}$ generated from a random latent code $\boldsymbol{z}$ (D: drums, B: bass, G: guitar, S: strings). Blue entries indicate the activation of single instruments at specific timesteps. Beats are numbered on the horizontal axis. (\subref{fig:editedstruct}) The edited structure tensor $\mathbf{\hat{S}}$. Green entries indicate the addition of new activations.}
\end{figure}

The separation of structure and content in our approach allows for the replacement of the generated structure tensor $\bold{S}$ with a new tensor $\Hat{\bold{S}}$ during the decoding process. This new tensor can be modified in a similar fashion to pianoroll editing in Digital Audio Workstations (DAW). For instance, the user can specify that a certain instrument should only be played at a specific time in the sequence by filling the desired positions in the binary activation grid. To show this, we operate as follows, focusing on the 2-bar model trained on the Lahk MIDI Dataset. We start by sampling a random latent code $\boldsymbol{z}$, from which we obtain the two representations $\boldsymbol{z}_\mathcal{S}$ and $\boldsymbol{z}_\mathcal{C}$. We then let the structure decoder produce the corresponding structure tensor $\bold{S}$ from $\boldsymbol{z}_\mathcal{S}$. At this point, we modify $\bold{S}$ to our liking, obtaining a new structure tensor $\Hat{\bold{S}}$. This corresponds to adding or removing nodes from the chord-level graph being generated. Finally, we let the content decoder compute two separate content tensors $\bold{X}$ and $\Hat{\bold{X}}$, corresponding to two final music sequences. For our purposes, the content decoder should be robust to changes in the structure, replicating the same musical content represented by $\boldsymbol{z}_\mathcal{C}$. When listening to the audio samples generated in this way, the model appears to be able to preserve the rhythmic and tonal features of the original sequence, rearranging the musical content while abiding by the imposed structure. As an example, Figure \ref{fig:genstruct} shows a generated structure tensor $\bold{S}$. The resulting sequence contains a recognizable I-IV progression in the key of B, supported by 8-beat bass and drum patterns\footnote{This and other examples related to conditioned generation can be found here: \url{https://emanuelecosenza.github.io/polyphemus/conditioned-generation.html}}. We edit the tensor by making the drums sparser, keeping only the nodes at the start of each beat, and by making the strings more active, adding new nodes at the start of beats. This yields a new structure tensor $\Hat{\bold{S}}$, which is shown in Figure \ref{fig:editedstruct}. The resulting music produced by the content decoder maintains the same harmonic progression of the original sequence. The bass and guitar tracks remain unaltered with very slight variations. Finally, the strings play a new melodic line in the right key, while the drums play a steady 4-beat hi-hat pattern. Overall, this shows that the content decoder can adapt to new structures specified by the user, opening the door to a new form of human-computer music co-creation.

\subsection{Embedding Visualization}

\begin{figure}
    \centering
    \begin{subfigure}{0.235\textwidth}
        \centering
        \includegraphics[width=1\textwidth]{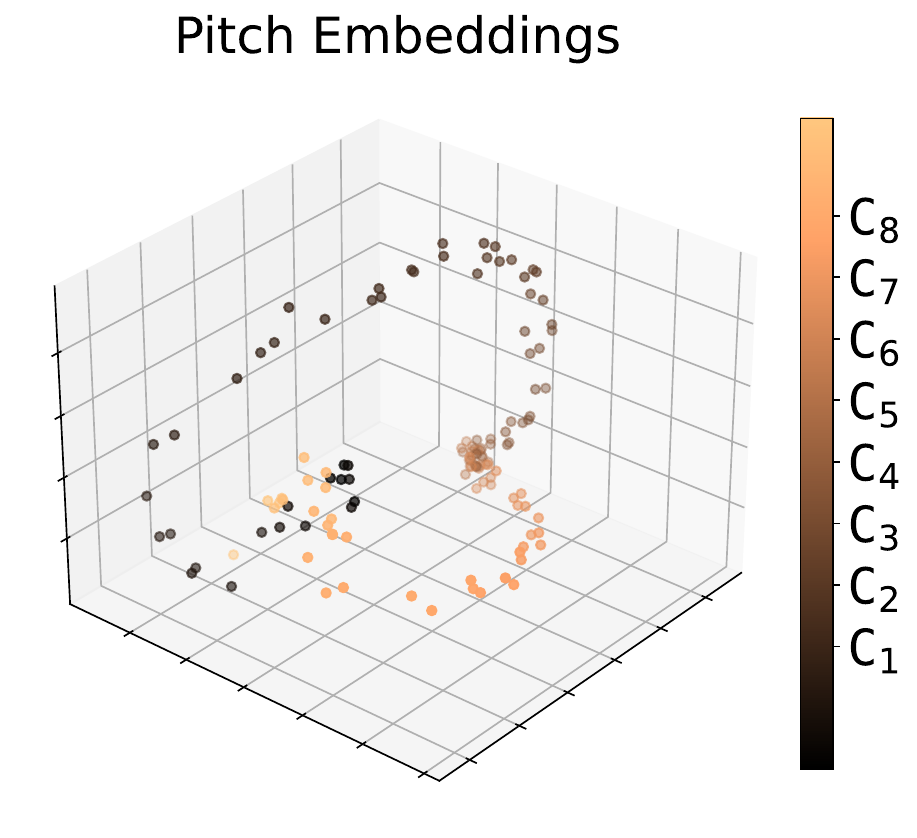}
        \caption{\label{fig:pitchembs}}
    \end{subfigure}
    \begin{subfigure}{0.235\textwidth}
        \centering
        \includegraphics[width=1\textwidth]{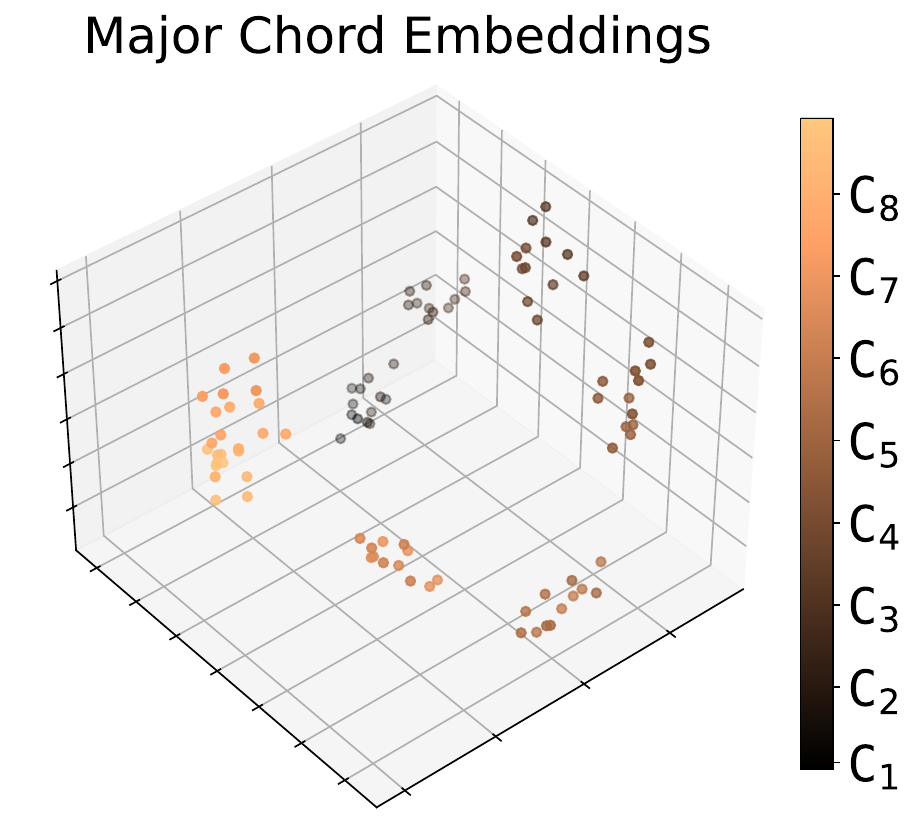}
        \caption{\label{fig:chordembs}}
    \end{subfigure}
\caption{(\subref{fig:pitchembs}) PCA projection of pitch embeddings. (\subref{fig:chordembs}) PCA projection of major chord embeddings. The major chords are obtained by picking as roots each note between $\texttt{C}_1$ and $\texttt{B}_8$.}
\label{fig:pitchchords}
\end{figure}

Similarly to \cite{wang2020pianotree} we explore the pitch, duration and chord embeddings by visualizing their principal components, focusing on the encoder network of the 2-bar model trained on the Lahk MIDI Dataset. Figure \ref{fig:pitchembs} shows the PCA projection in 3D space of all the 128 pitch embeddings. Pitch projections follow a circular path along the clockwise direction, suggesting that the model has learned the tonal relationships between different pitches. Figure \ref{fig:chordembs} shows a 3D PCA projection of chord embeddings considering every major chord obtained by picking as roots the notes between $\texttt{C}_1$ and $\texttt{B}_8$. Durations are fixed to 1 beat. Similarly to what happens for pitches, chord embeddings follow a circular path in the space and form clusters related to specific octaves.

\begin{figure}
    \centering
    \begin{subfigure}{0.235\textwidth}
        \centering
        \includegraphics[width=1\textwidth]{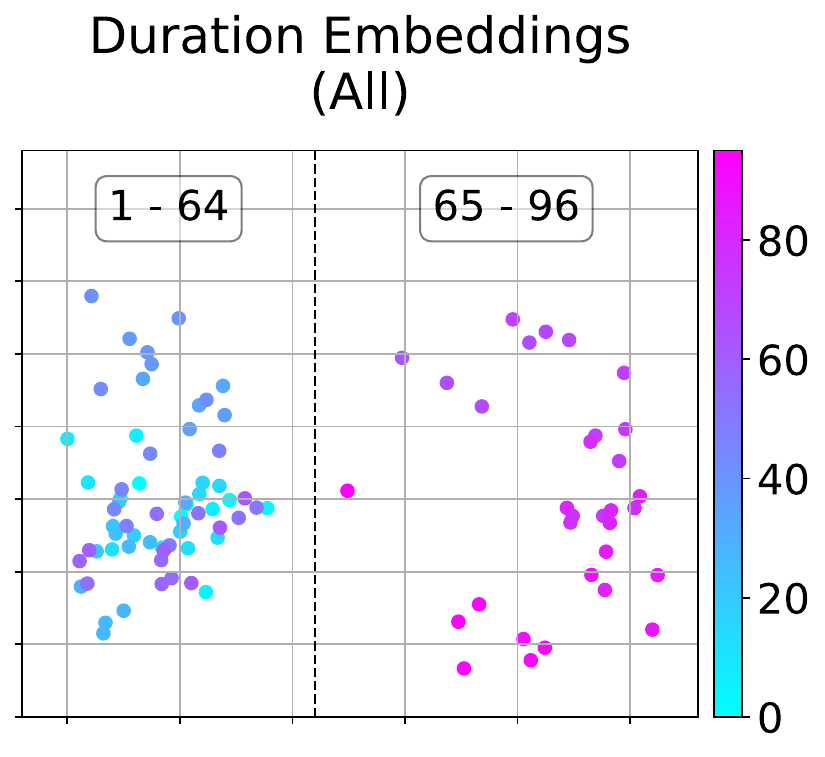}
        \caption{\label{fig:dur96}}
    \end{subfigure}
    \begin{subfigure}{0.235\textwidth}
        \centering
        \includegraphics[width=1\textwidth]{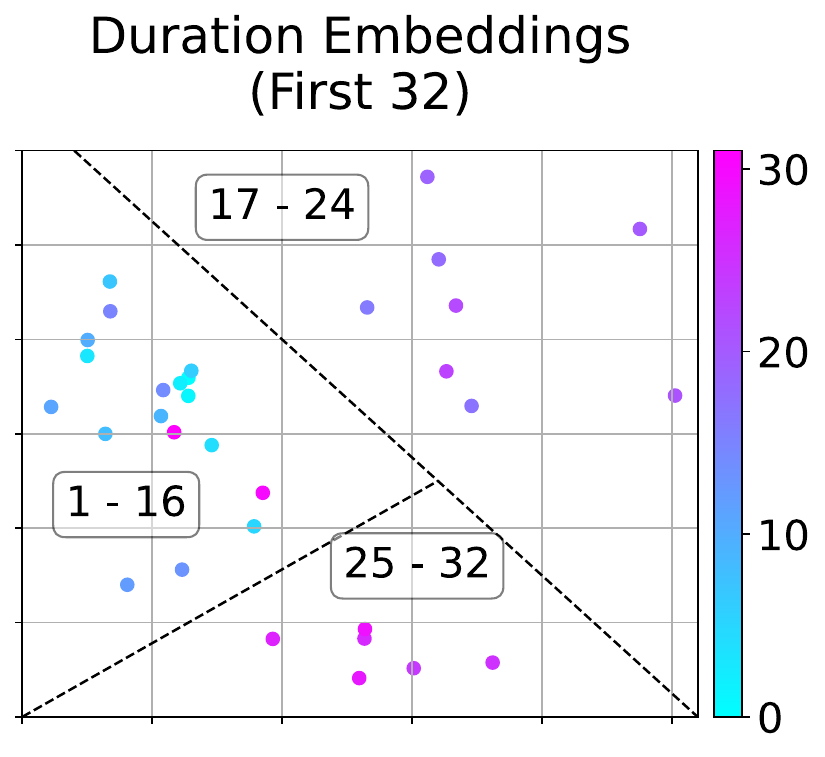}
        \caption{\label{fig:dur32}}
    \end{subfigure}
\caption{2D PCA projections of duration embeddings, considering all 96 durations (\subref{fig:dur96}) and the first 32 (\subref{fig:dur32}).}
\label{fig:durs}
\end{figure}

Figure \ref{fig:durs} shows the PCA projections in 2D space of duration embeddings considering, respectively, all the possible 96 durations (i.e. up to three bars) and the first 32 durations (i.e. up to a bar). In the first case (Figure \ref{fig:dur96}), two distinct clusters contain, respectively, durations above 64 (i.e. above 2 bars) and durations below 64 (i.e. below 2 bars). In the second plot (Figure \ref{fig:dur32}), three clusters can be identified with, respectively, durations below 16 (i.e. below 2 beats, left of the plot), durations between 16 and 24 (i.e. between 2 and 3 beats, upper-right of the plot) and durations above 24 (i.e. between 3 beats and a bar). The plots suggest that the model has learned to organize its duration space in accordance to the rhythmic concepts of beats and bars.

\section{Conclusions}

In this work, we introduced a new graph representation for polyphonic multitrack music and a model that generates musical graphs by separating their structure and content. We then tested our methodology on both known and unexplored MIDI datasets, considering short and long sequences. As seen in the qualitative analysis and the comparison with the state of the art, our approach has revealed to be beneficial with regards to the rhythmic and tonal consistency of the generated music. Through manual experiments, we showed that the models are able to replicate the same musical content when varying the structure of the graphs. This allows for a new generative scenario where users can specify the activity of particular instruments in a music sequence. To conclude, we further validated our methodology by visualizing the pitch, chord and duration embeddings learned by the model. In each case, the embedding spaces are organized in accordance with known tonal and rhythmic concepts.

This work represents a first attempt at generating music with graph-based deep methodologies and should be considered as a starting point for further research on the topic.
In the future, we aim to extend our work by taking into account MIDI velocity values, by training our model on other datasets and by studying new feasible graph representations and model configurations. Recurrent networks may be tested in the bar compressors and decompressors to check for improvements in the variability across bars. Again, a new \textit{sustain} edge type could model the sustain of notes in different track activations. To conclude, we believe that the model has the potential to support human-computer co-creation, and it will be interesting to find possible applications of our methodology in modern software audio tools.

\newpage

\bibliographystyle{named}
\bibliography{ijcai23}

\begin{thebibliography}{}

\bibitem[\protect\citeauthoryear{Agostinelli \bgroup \em et al.\egroup
  }{2023}]{agostinelli2023musiclm}
Andrea Agostinelli, Timo~I Denk, Zal{\'a}n Borsos, Jesse Engel, Mauro Verzetti,
  Antoine Caillon, Qingqing Huang, Aren Jansen, Adam Roberts, Marco
  Tagliasacchi, et~al.
\newblock Musiclm: Generating music from text.
\newblock {\em arXiv preprint arXiv:2301.11325}, 2023.

\bibitem[\protect\citeauthoryear{Bacciu \bgroup \em et al.\egroup
  }{2020}]{bacciu2020gentle}
Davide Bacciu, Federico Errica, Alessio Micheli, and Marco Podda.
\newblock A gentle introduction to deep learning for graphs.
\newblock {\em Neural Networks}, 129:203--221, 2020.

\bibitem[\protect\citeauthoryear{Brown \bgroup \em et al.\egroup
  }{2020}]{brown2020language}
Tom Brown, Benjamin Mann, Nick Ryder, Melanie Subbiah, Jared~D Kaplan, Prafulla
  Dhariwal, Arvind Neelakantan, Pranav Shyam, Girish Sastry, Amanda Askell,
  et~al.
\newblock Language models are few-shot learners.
\newblock {\em Advances in neural information processing systems},
  33:1877--1901, 2020.

\bibitem[\protect\citeauthoryear{Brunner \bgroup \em et al.\egroup
  }{2017}]{brunner2017jambot}
Gino Brunner, Yuyi Wang, Roger Wattenhofer, and Jonas Wiesendanger.
\newblock Jambot: Music theory aware chord based generation of polyphonic music
  with lstms.
\newblock In {\em 2017 IEEE 29th International Conference on Tools with
  Artificial Intelligence (ICTAI)}, pages 519--526. IEEE, 2017.

\bibitem[\protect\citeauthoryear{Brunner \bgroup \em et al.\egroup
  }{2018}]{brunner2018midi}
Gino Brunner, Andres Konrad, Yuyi Wang, and Roger Wattenhofer.
\newblock Midi-vae: Modeling dynamics and instrumentation of music with
  applications to style transfer.
\newblock {\em arXiv preprint arXiv:1809.07600}, 2018.

\bibitem[\protect\citeauthoryear{Chu \bgroup \em et al.\egroup
  }{2016}]{chu2016song}
Hang Chu, Raquel Urtasun, and Sanja Fidler.
\newblock Song from pi: A musically plausible network for pop music generation.
\newblock {\em arXiv preprint arXiv:1611.03477}, 2016.

\bibitem[\protect\citeauthoryear{Chuan and Herremans}{2018}]{chuan2018modeling}
Ching-Hua Chuan and Dorien Herremans.
\newblock Modeling temporal tonal relations in polyphonic music through deep
  networks with a novel image-based representation.
\newblock In {\em Proceedings of the AAAI Conference on Artificial
  Intelligence}, volume~32, 2018.

\bibitem[\protect\citeauthoryear{Dong \bgroup \em et al.\egroup
  }{2018}]{dong2018musegan}
Hao-Wen Dong, Wen-Yi Hsiao, Li-Chia Yang, and Yi-Hsuan Yang.
\newblock Musegan: Multi-track sequential generative adversarial networks for
  symbolic music generation and accompaniment.
\newblock In {\em Proceedings of the AAAI Conference on Artificial
  Intelligence}, volume~32, 2018.

\bibitem[\protect\citeauthoryear{Ens and Pasquier}{2021}]{ens2021building}
Jeffrey Ens and Philippe Pasquier.
\newblock Building the metamidi dataset: Linking symbolic and audio musical
  data.
\newblock In {\em ISMIR}, pages 182--188, 2021.

\bibitem[\protect\citeauthoryear{Ferretti}{2017}]{ferretti2017modeling}
Stefano Ferretti.
\newblock On the modeling of musical solos as complex networks.
\newblock {\em Information Sciences}, 375:271--295, 2017.

\bibitem[\protect\citeauthoryear{Ferretti}{2018}]{ferretti2018complex}
Stefano Ferretti.
\newblock On the complex network structure of musical pieces: analysis of some
  use cases from different music genres.
\newblock {\em Multimedia Tools and Applications}, 77(13):16003--16029, 2018.

\bibitem[\protect\citeauthoryear{Gilmer \bgroup \em et al.\egroup
  }{2017}]{gilmer2017neural}
Justin Gilmer, Samuel~S Schoenholz, Patrick~F Riley, Oriol Vinyals, and
  George~E Dahl.
\newblock Neural message passing for quantum chemistry.
\newblock In {\em International conference on machine learning}, pages
  1263--1272. PMLR, 2017.

\bibitem[\protect\citeauthoryear{Goodfellow \bgroup \em et al.\egroup
  }{2014}]{goodfellow2014generative}
Ian Goodfellow, Jean Pouget-Abadie, Mehdi Mirza, Bing Xu, David Warde-Farley,
  Sherjil Ozair, Aaron Courville, and Yoshua Bengio.
\newblock Generative adversarial nets.
\newblock {\em Advances in neural information processing systems}, 27, 2014.

\bibitem[\protect\citeauthoryear{Goodfellow \bgroup \em et al.\egroup
  }{2016}]{Goodfellow-et-al-2016}
Ian Goodfellow, Yoshua Bengio, and Aaron Courville.
\newblock {\em Deep Learning}.
\newblock MIT Press, 2016.
\newblock \url{http://www.deeplearningbook.org}.

\bibitem[\protect\citeauthoryear{Higgins \bgroup \em et al.\egroup
  }{2016}]{higgins2016beta}
Irina Higgins, Loic Matthey, Arka Pal, Christopher Burgess, Xavier Glorot,
  Matthew Botvinick, Shakir Mohamed, and Alexander Lerchner.
\newblock beta-vae: Learning basic visual concepts with a constrained
  variational framework.
\newblock 2016.

\bibitem[\protect\citeauthoryear{Huang \bgroup \em et al.\egroup
  }{2018}]{huang2018music}
Cheng-Zhi~Anna Huang, Ashish Vaswani, Jakob Uszkoreit, Noam Shazeer, Ian Simon,
  Curtis Hawthorne, Andrew~M Dai, Matthew~D Hoffman, Monica Dinculescu, and
  Douglas Eck.
\newblock Music transformer.
\newblock {\em arXiv preprint arXiv:1809.04281}, 2018.

\bibitem[\protect\citeauthoryear{Huang \bgroup \em et al.\egroup
  }{2019}]{huang2019counterpoint}
Cheng-Zhi~Anna Huang, Tim Cooijmans, Adam Roberts, Aaron Courville, and Douglas
  Eck.
\newblock Counterpoint by convolution.
\newblock {\em arXiv preprint arXiv:1903.07227}, 2019.

\bibitem[\protect\citeauthoryear{Ioffe and Szegedy}{2015}]{ioffe2015batch}
Sergey Ioffe and Christian Szegedy.
\newblock Batch normalization: Accelerating deep network training by reducing
  internal covariate shift.
\newblock In {\em International conference on machine learning}, pages
  448--456. PMLR, 2015.

\bibitem[\protect\citeauthoryear{Jeong \bgroup \em et al.\egroup
  }{2019}]{jeong2019graph}
Dasaem Jeong, Taegyun Kwon, Yoojin Kim, and Juhan Nam.
\newblock Graph neural network for music score data and modeling expressive
  piano performance.
\newblock In {\em International Conference on Machine Learning}, pages
  3060--3070. PMLR, 2019.

\bibitem[\protect\citeauthoryear{Karystinaios and
  Widmer}{2022}]{karystinaios2022cadence}
Emmanouil Karystinaios and Gerhard Widmer.
\newblock Cadence detection in symbolic classical music using graph neural
  networks, 2022.

\bibitem[\protect\citeauthoryear{Kingma and Ba}{2014}]{kingma2014adam}
Diederik~P Kingma and Jimmy Ba.
\newblock Adam: A method for stochastic optimization.
\newblock {\em arXiv preprint arXiv:1412.6980}, 2014.

\bibitem[\protect\citeauthoryear{Kingma and Welling}{2013}]{kingma2013auto}
Diederik~P Kingma and Max Welling.
\newblock Auto-encoding variational bayes.
\newblock {\em arXiv preprint arXiv:1312.6114}, 2013.

\bibitem[\protect\citeauthoryear{Li \bgroup \em et al.\egroup
  }{2015}]{li2015gated}
Yujia Li, Daniel Tarlow, Marc Brockschmidt, and Richard Zemel.
\newblock Gated graph sequence neural networks.
\newblock {\em arXiv preprint arXiv:1511.05493}, 2015.

\bibitem[\protect\citeauthoryear{Li \bgroup \em et al.\egroup
  }{2018}]{li2018deeper}
Qimai Li, Zhichao Han, and Xiao-Ming Wu.
\newblock Deeper insights into graph convolutional networks for semi-supervised
  learning.
\newblock In {\em Thirty-Second AAAI conference on artificial intelligence},
  2018.

\bibitem[\protect\citeauthoryear{Li \bgroup \em et al.\egroup
  }{2019}]{li2019deepgcns}
Guohao Li, Matthias Muller, Ali Thabet, and Bernard Ghanem.
\newblock Deepgcns: Can gcns go as deep as cnns?
\newblock In {\em Proceedings of the IEEE/CVF international conference on
  computer vision}, pages 9267--9276, 2019.

\bibitem[\protect\citeauthoryear{Liu \bgroup \em et al.\egroup
  }{2010}]{liu2010complex}
Xiao~Fan Liu, K~Tse Chi, and Michael Small.
\newblock Complex network structure of musical compositions: Algorithmic
  generation of appealing music.
\newblock {\em Physica A: Statistical Mechanics and its Applications},
  389(1):126--132, 2010.

\bibitem[\protect\citeauthoryear{Makhzani \bgroup \em et al.\egroup
  }{2015}]{makhzani2015adversarial}
Alireza Makhzani, Jonathon Shlens, Navdeep Jaitly, Ian Goodfellow, and Brendan
  Frey.
\newblock Adversarial autoencoders.
\newblock {\em arXiv preprint arXiv:1511.05644}, 2015.

\bibitem[\protect\citeauthoryear{Mellon \bgroup \em et al.\egroup
  }{2014}]{mellon2014genre}
Rachel Mellon, Dan Spaeth, and Eric Theis.
\newblock Genre classification using graph representations of music.
\newblock 2014.

\bibitem[\protect\citeauthoryear{Mogren}{2016}]{mogren2016c}
Olof Mogren.
\newblock C-rnn-gan: Continuous recurrent neural networks with adversarial
  training.
\newblock {\em arXiv preprint arXiv:1611.09904}, 2016.

\bibitem[\protect\citeauthoryear{Raffel}{2016}]{raffel2016learning}
Colin Raffel.
\newblock {\em Learning-based methods for comparing sequences, with
  applications to audio-to-midi alignment and matching}.
\newblock Columbia University, 2016.

\bibitem[\protect\citeauthoryear{Ramesh \bgroup \em et al.\egroup
  }{2022}]{ramesh2022hierarchical}
Aditya Ramesh, Prafulla Dhariwal, Alex Nichol, Casey Chu, and Mark Chen.
\newblock Hierarchical text-conditional image generation with clip latents.
\newblock {\em arXiv preprint arXiv:2204.06125}, 2022.

\bibitem[\protect\citeauthoryear{Roberts \bgroup \em et al.\egroup
  }{2018}]{roberts2018hierarchical}
Adam Roberts, Jesse Engel, Colin Raffel, Curtis Hawthorne, and Douglas Eck.
\newblock A hierarchical latent vector model for learning long-term structure
  in music.
\newblock In {\em International conference on machine learning}, pages
  4364--4373. PMLR, 2018.

\bibitem[\protect\citeauthoryear{Schlichtkrull \bgroup \em et al.\egroup
  }{2018}]{schlichtkrull2018modeling}
Michael Schlichtkrull, Thomas~N Kipf, Peter Bloem, Rianne van~den Berg, Ivan
  Titov, and Max Welling.
\newblock Modeling relational data with graph convolutional networks.
\newblock In {\em European semantic web conference}, pages 593--607. Springer,
  2018.

\bibitem[\protect\citeauthoryear{Simonetta \bgroup \em et al.\egroup
  }{2018}]{simonetta2018symbolic}
Federico Simonetta, Filippo Carnovalini, Nicola Orio, and Antonio Rod{\`a}.
\newblock Symbolic music similarity through a graph-based representation.
\newblock In {\em Proceedings of the Audio Mostly 2018 on Sound in Immersion
  and Emotion}, pages 1--7. 2018.

\bibitem[\protect\citeauthoryear{Simonovsky and
  Komodakis}{2017}]{simonovsky2017dynamic}
Martin Simonovsky and Nikos Komodakis.
\newblock Dynamic edge-conditioned filters in convolutional neural networks on
  graphs.
\newblock In {\em Proceedings of the IEEE conference on computer vision and
  pattern recognition}, pages 3693--3702, 2017.

\bibitem[\protect\citeauthoryear{Valenti \bgroup \em et al.\egroup
  }{2020}]{valenti2020learning}
Andrea Valenti, Antonio Carta, and Davide Bacciu.
\newblock Learning style-aware symbolic music representations by adversarial
  autoencoders.
\newblock {\em arXiv preprint arXiv:2001.05494}, 2020.

\bibitem[\protect\citeauthoryear{Valenti \bgroup \em et al.\egroup
  }{2021}]{valenti2021calliope}
Andrea Valenti, Stefano Berti, and Davide Bacciu.
\newblock Calliope--a polyphonic music transformer.
\newblock {\em arXiv preprint arXiv:2107.05546}, 2021.

\bibitem[\protect\citeauthoryear{Wang \bgroup \em et al.\egroup
  }{2020}]{wang2020pianotree}
Ziyu Wang, Yiyi Zhang, Yixiao Zhang, Junyan Jiang, Ruihan Yang, Junbo Zhao, and
  Gus Xia.
\newblock Pianotree vae: Structured representation learning for polyphonic
  music.
\newblock {\em arXiv preprint arXiv:2008.07118}, 2020.

\bibitem[\protect\citeauthoryear{Wu \bgroup \em et al.\egroup
  }{2020}]{wu2020popmnet}
Jian Wu, Xiaoguang Liu, Xiaolin Hu, and Jun Zhu.
\newblock Popmnet: Generating structured pop music melodies using neural
  networks.
\newblock {\em Artificial Intelligence}, 286:103303, 2020.

\bibitem[\protect\citeauthoryear{Yang \bgroup \em et al.\egroup
  }{2017}]{yang2017midinet}
Li-Chia Yang, Szu-Yu Chou, and Yi-Hsuan Yang.
\newblock Midinet: A convolutional generative adversarial network for
  symbolic-domain music generation.
\newblock {\em arXiv preprint arXiv:1703.10847}, 2017.

\bibitem[\protect\citeauthoryear{Zou \bgroup \em et al.\egroup
  }{2021}]{zou2021melons}
Yi~Zou, Pei Zou, Yi~Zhao, Kaixiang Zhang, Ran Zhang, and Xiaorui Wang.
\newblock Melons: generating melody with long-term structure using transformers
  and structure graph.
\newblock {\em arXiv preprint arXiv:2110.05020}, 2021.

\end{thebibliography}

\appendix

\end{document}